\documentclass{article}
            \usepackage{graphicx}
            \usepackage{amsmath,amssymb}
            \begin{document}
            \title{Local availability of mathematics and number scaling: Effects on  quantum physics}
            \author{Paul Benioff,\\
            Physics Division, Argonne National
            Laboratory,\\ Argonne, IL 60439, USA \\
            e-mail:pbenioff@anl.gov}

            \maketitle

            \begin{abstract}
            Local availability of mathematics and  number scaling provide an  approach to a coherent theory of physics and mathematics.  Local availability of mathematics  assigns separate mathematical universes, $\bigcup_{x},$ to each space time point, $x.$.  The mathematics available to an observer, $O_{x},$ at $x$ is  contained in $\bigcup_{x}.$ Number scaling is based on extending the  choice freedom of vector space bases in gauge theories to choice freedom of underlying number systems. Scaling arises in the description, in $\bigcup_{x},$ of  mathematical systems in $\bigcup_{y}.$ If $a_{y}$ or $\psi_{y}$  is a number or a quantum state in $\bigcup_{y}$, then the corresponding number or state in $\bigcup_{x}$ is $r_{y,x}a_{x}$ or $r_{y,x}\psi_{x}.$ Here $a_{x}$ and $\psi_{x}$ are the same number and state in $\bigcup_{x}$ as $a_{y}$ and $\psi_{y}$ are in $\bigcup_{y}.$ If $y=x+\hat{\mu}dx$ is a neighbor point of $x,$ then the scaling factor is $r_{y,x}=\exp(\vec{A}(x)\cdot\hat{\mu}dx)$ where $\vec{A}$ is the gradient of a scalar field.

            The effects of scaling and local availability of mathematics on quantum theory show that scaling has two components, external and internal. External scaling is shown above for $a_{y}$ and $\psi_{y}.$ Internal scaling occurs  in expressions with integrals or derivatives over space time. An example is the replacement of the position expectation value, $\int\psi^{*}(y)y\psi(y)dy,$ by $\int_{x}r_{y,x}\psi^{*}_{x}(y_{x})y_{x}\psi_{x}(y_{x})dy_{x}.$ This is an integral in $\bigcup_{x}.$

            The good agreement between quantum theory and experiment shows that scaling is negligible in a space region, $L$, in which  experiments and calculations can be done, and results compared. $L$ includes the solar system, but the speed of light limits the size of $L$ to a  few light years. For observers in $L$ and events outside $L,$ at cosmological distances,  scaling is not limited by theory experiment agreement requirements.
            \end{abstract}

            \section{Introduction}
            The nature of mathematics and its relation to physics has been and is a topic of much interest \cite{Hut,Tegmark,Bernal,Welch,Jannes}. Wigner's question "Why is mathematics relevant to the natural sciences?" \cite{Wigner} continues to be of much interest \cite{Omnes,Hamming}. It has led to  attempts by this author \cite{BenCTPM1,BenCTPM2} to work towards a coherent theory of mathematics and physics together which hopefully will help to answer this question.

            This paper extends work in this direction by  exploration of the effects of local availability of mathematics and number scaling on  a few aspects of quantum physics and geometry.  Since the local availability of mathematics and number scaling have been discussed in earlier work \cite{BenLAM}, the discussions here will be brief and cover the main points.

            Local availability of mathematics and number scaling are extensions of some basic aspects of gauge theories. In these theories one assigns an $n$ dimensional vector space $\bar{V}_{x}$ to each point, $x$, in space time \cite{Montvay}. Matter fields, $\psi,$ take values in these spaces where $\psi(x)\epsilon \bar{V}_{x}.$  Yang  and Mills \cite{Yang} introduced the freedom of choice of bases in these spaces in that the choice of bases on one space does not determine the bases in another space.

            In gauge theories, the  complex number scalars,  $\bar{C}$, are the same for all the vector spaces.  This restriction  can be removed by assignment of separate complex number structures, $\bar{C}_{x},$ to each space time point, $x$ \cite{BenLAM}. The freedom of choice of bases in the different vector spaces is extended to the freedom of choice of number values in the different $\bar{C}_{x}$.

            The freedom of choice of number values is implemented here by use of a real valued scalar field, $\theta,$  such that for each $x,$ $\theta(x)$ is a number in $\bar{C}_{x}.$  This field is used to define the relations between number values in the different $\bar{C}_{x}$ in a manner similar to the use of elements of the gauge group to define relations between  bases in the different $\bar{V}_{x}$ in gauge theories \cite{Montvay,Cheng}. Here the relations between the number values in the different $\bar{C}_{x}$ are defined  by  real scaling factors.

            These  factors are defined from a vector field, $\vec{A},$ that is the gradient of $\theta.$  If $y=x+\hat{\mu}dx$ then the scaling factor, $r_{y,x}$, that relates $\bar{C}_{y}$ to $\bar{C}_{x}$ is defined by\begin{equation} \label{defryx}r_{y,x}=e^{\vec{A}(x)\cdot\hat{\mu}dx}\end{equation} where
            \begin{equation}\label{Agrth}\vec{A}(x)=\nabla_{x}\theta.\end{equation}If $y$ is distant from $x$ then, according to the gradient theorem, \cite{VFWiki}, $r_{y,x}$ is independent of the path from $x$ to $y$ and \begin{equation}\label{rgdth}r_{y,x}=e^{\theta(y)-\theta(x)}.\end{equation}Here $r_{y,x}$ is a real number value in $\bar{C}_{x}.$

            As is well known, numbers are an important component of many different mathematical systems.  Included are various spaces, algebras, group representations, etc. This suggests that one extend the assignment of separate $\bar{C}_{x}$ to each $x,$ to any mathematical system that includes numbers in its definition.\footnote{The scaling defined here for the complex numbers also applies to all the other number types, especially if they are considered as subsets of the complex numbers.} If $\bar{S}$ is such a system, then $\bar{S}_{x}$ is the system at point $x$ just as $\bar{C}_{x}$ is the complex number system at $x.$ Scaling  factors affecting the relations between the different complex number fields would be expected to have an effect on the relations between the different $\bar{S}$ type systems.

            The many different mathematical systems whose definitions depend at least partly on numbers can be collected into mathematical universes, $\bigcup_{x}$, one for each point $x.$  Each $\bigcup_{x}$ contains all types of systems $\bar{S}_{x}$ whose axiomatic definitions include numbers of some type in their definitions.  The assignment of separate universes to each point, $x,$ is supported by the concept of the local availability of mathematics. This, and the contents of the different $\bigcup_{x},$ and their relation to one another at different locations are discussed in the next section. It is seen that scaling does not have an effect on comparison of  theoretical computations and experimental outputs.

            Number scaling is discussed in Section \ref{NS}. The mathematical logical definition of mathematical systems, as structures \cite{Barwise, Keisler}, is used. A structure consists of a set, basic operations, relations, and a few constants where the structure satisfies a set of axioms relevant to the system type. It is seen that it is possible to define structures of each number type that differ by arbitrary scaling factors and satisfy the relevant axioms if and only if a structure without scaling satisfies the axioms. In these structures, number value scaling is compensated for by scaling of  some of the basic operations in the structures.

            Section \ref{SFBCC} applies number scaling to the  complex number structures, $\bar{C}_{x}$ at each point $x.$ It is seen that, for an observer, $O_{x},$ at point $x,$ the local representation of the complex number structure, $\bar{C}_{y},$ on $\bar{C}_{x}$, includes a scaling factor, $r_{y,x}$.  For $y$ distant from $x$ $r_{y,x}$ is given by Eq. \ref{rgdth}. Here $\bar{C}_{x}$ is the complex number base for mathematics available to $O_{x}.$

            The work of these sections is applied in section \ref{QT} to quantum theory. It is seen that there are two types of scaling, external and internal. External scaling is used to map the mathematical representations of system properties at $y,$  to another point $x.$ These are the  representations, in $\bigcup_{x},$ at $x,$ of  descriptions of the system properties based on mathematics in $\bigcup_{y}$ at another point, $y.$

            Internal scaling applies only to properties of systems expressed as integrals or derivatives over space or space time.  The integrands, associated with each point in space, are taken to belong to the universes associated with the point. In order for the integral to make sense, the integrands must be transferred, with scaling to a common reference point where the integral can be defined. For derivatives, such as the momentum operator the usual momentum is replaced by the canonical momentum that includes the $\vec{A}$ field, Eq. \ref{defryx}.

            The next two sections describe space regions in which scaling is negligible, and in which there are, no theory experiment restrictions on scaling.  The region, $L$, in which scaling must be negligible, includes locations in which theory calculations and experiments can be implemented. Here $L$ is assumed to be a region that includes the solar system and is a few lightyears in size. For points $y$ outside $L$, which is most of the cosmos, there are no theory experiment restrictions on scaling involved in the representation, at any point $x$ in $L$, of properties of systems at $y.$

            The concludes with a discussion. It includes a brief review and an emphasis of the amount of work yet to be done.
            \section{local Availability of Mathematics}\label{LAM}
            As was noted, the totality of mathematics available to an observer, $O_{x},$ at point $x$ is contained in $\bigcup_{x}.$ That is, mathematics is available locally. This is supported by the observation that all mathematics (or physics) that $O_{x}$ knows or is aware of is in his or her brain. Knowledge of  new mathematics is obtained only when  the requisite mathematical information in a book or in a seminar talk is transmitted to the observer by sound or light and is received in the brain. For the purposes of this work details of this process are irrelevant.

            Local availability of mathematics extends to observers motion in space time.  This is described by a world line, $w(\tau)$ where $\tau$ denotes the proper time. The mathematics available to $O_{w(\tau)}$ at proper time $\tau$ is that contained in $\bigcup_{w(\tau)}.$

            The contents of each $\bigcup_{x}$ are based on the mathematical logical definitions of mathematical systems as structures \cite{Barwise,Keisler}. A structure consists of a base set of elements, a few basic operations, basic relations, and constants that satisfy a set of axioms relevant to the type of system being considered. As examples, a rational number structure, $\overline{Ra}_{x}$ in $\bigcup_{x}$ is defined as a structure, \begin{equation}\label{Ra}\overline{Ra}_{x}=\{Ra_{x},+_{x},-_{x}, \times_{x},\div_{x},<_{x}, 0_{x},1_{x}\}\end{equation} that satisfies the axioms for the smallest ordered field \cite{rational}. Structures for the real and complex numbers are given by \begin{equation}\label{Real}\overline{R}_{x}=\{R_{x},+_{x}, -_{x}, \times_{x},\div_{x},<_{x}, 0_{x},1_{x}\}\end{equation}and
            \begin{equation}\label{Complex}\bar{C}_{x}=\{C_{x},+_{x},-_{x}, \times_{x},\div_{x}, <_{x}, 0_{x},1_{x}\}.\end{equation} These structures satisfy, respectively, axioms for a complete ordered field \cite{real} and an algebraically closed field of characteristic $0$ \cite{complex}. A Hilbert space structure,
            \begin{equation}\label{Hilbert}\bar{H}_{x}=\{H_{x},+_{x},-_{x},\cdot_{x}, \langle-,-\rangle_{x},\psi\}\end{equation} satisfies the axioms for a complex, normed, inner product space that is complete in a norm defined from the inner product, \cite{Kadison}.

            In these structures, a symbol with an overline, such as $\bar{R},$ denotes a structure.  The same symbol with no overline, such as $R$ denotes a base set for the structure.  The subscript $x$, denotes structure membership in $\bigcup_{x}.$ For Hilbert spaces, $\cdot_{x}$ and $\langle-,-\rangle_{x}$ denote scalar vector multiplication and scalar or inner product, and $\psi$ denotes any state.

            The universes are all equivalent in that if $\bar{S}_{x}$ is a structure in $\bigcup_{x}$, then there is a structure, $\bar{S}_{y},$ in $\bigcup_{y}$ that is the same structure in $\bigcup_{y}$ as $\bar{S}_{x}$ is in $\bigcup_{x}$.

            The mathematics available to $O_{x}$ must include the ability to describe properties of physical systems at different locations. Integrals and derivatives over space time must be possible to describe in $\bigcup_{x}.$ This requires a map of structures in $\bigcup_{y}$ onto structures in $\bigcup_{x}.$

            A basic map is a parallel transformation \cite{Mack}. This map  defines the notion of sameness between elements of structures  in $\bigcup_{y}$ and $\bigcup_{x}.$ Let $\mathcal{F}_{y,x}:\bar{S}_{x}\rightarrow\bar{S}_{y}$   be a parallel transformation from  $\bar{S}_{x}$ to $\bar{S}_{y}.$ If $\bar{S}_{x}=\{S_{x},Op_{x},Re_{x},Co_{x}\}$ and $\bar{S}_{y}=\{S_{y},Op_{y} ,Re_{y},Co_{y}\},$ then\begin{equation}\label{defFyx}
            \begin{array}{c}\mathcal{F}_{y,x}(S_{x})=S_{y},\hspace{1cm} \mathcal{F}_{y,x}(Op_{x})=Op_{y}, \\ \mathcal{F}_{y,x}(Re_{x})=Re_{y}, \hspace{1cm}\mathcal{F}_{y,x}(Co_{x})=Co_{y}.\end{array}\end{equation} Here $Op,Re,Co$ denote, collectively, the basic operations, relations, and constants of $\bar{S}.$ $\mathcal{F}_{y,x}$ maps $S_{x}$ elementwise onto $S_{y}$ such that for each element, $s_{x}$ in $S_{x}$, $s_{y}=\mathcal{F}_{y,x}(s_{x})$ is an element in $S_{y}$ that has the same value, relative to $\bar{S}_{y},$ as $s_{x}$ has in $\bar{S}_{x}.$ The map $\mathcal{F}_{y,x}$ can be easily extended to structures that include other structures in their definitions. An example is a Hilbert space structure, $\bar{H}_{x},$  that includes  a complex number structure $\bar{C}_{x},$ in its characterization. Also note that $\mathcal{F}_{y,x}$ has an inverse, denoted by $\mathcal{F}_{x,y}.$

            The next step is to introduce scaling between number structures at different points. Before doing this it is useful to show that what appears to be an obvious criticism does not apply. Assume that an experiment or computation is carried out at $x$ and that the same experiment or computation is also carried out at $y.$ Let $a_{x}$ and $b_{y}$ be the real numerical outcomes at $x$ and $y.$

            Comparison of the two outcomes requires mapping $a_{x}$ and $b_{y}$ to a common point so they can be compared within the same number structure.  Since the computations or experiments are the same at $x$ and $y,$ one expects $b_{y}$ to be the same number value in $\bar{C}_{y}$ as $a_{x}$ is in $\bar{C}_{x}.$ This is indeed the case if $a_{x}=\mathcal{F}_{x,y}(b_{y})$ which says that $a_{x}$ is the same number in $\bar{C}_{x}$ as $b_{y}$ is in $\bar{C}_{y}.$ (Statistics and quantum uncertainties are ignored here.)

            This is not the case if scaling is present as then the outcome $b_{y}$ corresponds to the number $c_{x}=r_{y,x}\mathcal{F}_{x,y}(b_{y}).$ Since $c_{x}\neq a_{x}$ if $r_{y,x}\neq 1$, which is the case if scaling is present, there is a problem.  In particular, such comparisons between theory and experiment, or between different computations or experiments, give no hint of the presence of scaling.

            This problem is resolved by noting that scaling plays no role in such comparisons. This is based on the observation that no computation and no experiment ever gives directly a number as an outcome. Instead, outcomes of of experiments or computations are physical systems in physical states that are \emph{interpreted} as numbers. Examples include strings of symbols on paper or computer screens, pointer positions, etc.

            To see how this works, let $\psi_{y}$ and $\phi_{x}$ be outputs of experiments at $x$ and $y.$ Then the real numbers in $\bar{C}_{y}$ and $\bar{C}_{x}$ corresponding to these outputs are given by $I_{y}(\psi_{y})$ and $I_{x}(\phi_{x}).$ For each $x,$ $I_{x}$ is a map from the set of physical output states of experiments or computations at $x$ to real number values in $\bar{C}_{x}.$

            The "naheinformationsprinzip" (no information at a distance) \cite{Montvay,Mack} principle forbids direct comparison of the information contained in $\psi_{y}$ and $\phi_{x}.$  Instead the information in these states must be transported by physical means to a common point, $z,$ for comparison. The method of transmission, by use of sound waves, optical transmission, etc. must be such that the information is preserved during transmission.

\begin{figure}[h!]\begin{center}
            \rotatebox{270}{\resizebox{140pt}{140pt}{\includegraphics[100pt,130pt]
            [520pt,570pt]{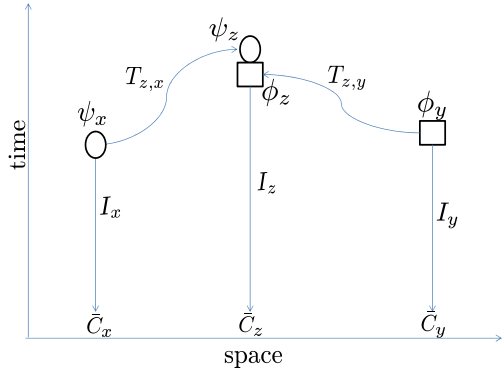}}}\end{center}
            \caption{A simple example of comparing theory with experiment.   The oval and square at locations $x,y$ denote the output computation and experiment systems in states $\psi_{x}$ and $\phi_{y}$. The $T_{z,x}$ and $T_{z,y}$ denote motion of these output states to a common point $z,$ as $\psi_{z}=T_{z,x}(\psi_{x})$ and $\phi_{z}=T_{z,y}(\phi_{y})$ where they can be compared. The $I_{x},I_{y},I_{z}$ denote interpretations of these states as numbers at locations $x,y,z.$ }\label{SPIE21}\end{figure}

            Let $T_{z,x}(\psi_{x})$ and $T_{z,y}(\phi_{y})$ denote these transmissions to $z$. Now the numerical values for the outcomes can be directly compared as they are both numbers in $\bar{C}_{z}.$ These values are given by $I_{z}(T_{z,x}(\psi_{x}))$ and $I_{z}(T_{z,y}(\phi_{y})).$ This process is shown schematically in Figure \ref{SPIE21}.\\

            Note that scaling plays no role in this comparison of outcomes. It follows that in regions of space time in which theoretical calculations can be done and experiments can be carried out, the effect of scaling, if any, must be below the accuracy of the theory experiment comparison.  However this does not mean that scaling must be negligible in all of space time. As will be seen later, scaling may be present in theoretical descriptions of events at cosmological distances.

            \section{Number Scaling}\label{NS}
            Number scaling arises from the observation that, for each type of number, it is possible to define many different number structures that differ by arbitrary scaling factors. For complex numbers, let $\bar{C}^{r}$ denote a complex number structure, \begin{equation}\label{Cr}\bar{C}^{r}=\{C,+_{r},-_{r}, \times_{r},\div_{r},0_{r},1_{r}\}.\end{equation} The super and subscripts $r$ show that this structure is scaled by a factor $r$, relative to $\bar{C},$ Eq. \ref{Complex}.

            The scaling can be shown explicitly in  a representation of $\bar{C}^{r}$ on $\bar{C}.$ This representation defines the structure elements and operations of $\bar{C}^{r}$ in terms of those in $\bar{C}.$  The result is\begin{equation}\label{CrC}\bar{C}^{r}=\{C,+,-, \frac{\times}{r}, r\div,0,r\}.\end{equation} This shows that $+$ and $-$ operations are unchanged by scaling, $\times_{r}$ in $\bar{C}^{r}$ corresponds to $\times/r$ in $\bar{C},$   $\div_{r}$ corresponds to $r\div$ in $\bar{C},$ and $0_{r}$ and $1_{r}$ correspond to $0$ and $r.$ Also any number value $a_{r}$ in $\bar{C}^{r}$ corresponds to the number value $ra$ in $\bar{C}.$ The number value $0$ is the only value unchanged by scaling.  In this sense it corresponds to the "number vacuum".

            The correspondences are defined to satisfy a basic requirement: that $\bar{C}^{r}$ satisfies the axioms for complex numbers if and only if $\bar{C}$ satisfies the axioms. As an example it can seen, from the axiom for multiplicative identity, that the number value $r$ in $\bar{C},$ when viewed in $\bar{C}^{r},$ is the the number value $1.$  This follows from the equivalences \begin{equation}\label{idax}a_{r}\times_{r}1_{r}=a_{r}\Leftrightarrow ra\frac{\times}{r}r=ra\Leftrightarrow a\times 1=a.\end{equation}The leftmost equation is defined on $\bar{C}^{r}$ and the other two are defined on $\bar{C}.$

            Note that scaling introduces a new relation, that of correspondence. This is different from the concept of "same value as".  For example the number value $a_{r}$ in $\bar{C}^{r}$ corresponds to the number value $ra$ in $\bar{C}$. This is different from $a$ which is the same number value in $\bar{C}$ as $a_{r}$ is in $\bar{C}^{r}.$ These two concepts, correspondence and same value as, coincide if $r=1.$ This is the usual case with no scaling present.

            The relations between the two structures show clearly that elements of the base set, $C,$  which is the same in both representations, have no fixed values.  The values associated with the elements are dependent on the structure containing $C.$  For example, the element, $s,$ of $C$ that has value $a_{r}$ in $\bar{C}^{r}$ has value $ra$ in $\bar{C}.$

            This is why numbers in structures are referred to as number values. If $s$ is called a number, then the values, $a_{r}$ or $ra,$ associated with $s,$ depend on the structure containing $s.$

            The relations between $\bar{C}^{r}$ and $\bar{C}$ show an interesting property. To see this let $f_{r}:\bar{C}^{r}\rightarrow\bar{C}^{r}$ be an analytic function on $\bar{C}^{r}.$  The equivalences, \begin{equation}\label{frar}
            f_{r}(a_{r})=c_{r}\Leftrightarrow rf(a)=rc\Leftrightarrow f(a)=c\end{equation} show that the function, $f_{r}$ on $\bar{C}_{r}$ that corresponds to $f$ on $\bar{C},$ is also the same function on $\bar{C}^{r}$ as $f$ is on $\bar{C}.$

            This follows from the observation that, for any term, $a_{r}^{m}/b_{r}^{n},$ the correspondence \begin{equation}\label{arm}\frac{a_{r}^{m}} {b_{r}^{n}}\mbox{}_{r}\rightarrow r\frac{a^{m}}{b^{n}}\end{equation}holds. The m factors, $a_{r}$ contribute a factor of $r^{m}$. The $m-1$ multiplications contribute a factor of $r^{-(m-1)}$ to give a net factor of $r.$  This is canceled by a similar $r$ factor from the denominator, $b_{r}^{m}.$ The final $r$ factor comes from the relation between division in $\bar{C}^{r}$ and $\bar{C}.$ More details on number scaling are provided in \cite{BenNS}.

            \section{Scaling Factors between $\bar{C}_{x}$ and $\bar{C}_{y}.$}\label{SFBCC}
            The freedom to choose scaling between the complex number structures (and structures of other number types) in the $\bigcup$ universes can be expressed by use of the parallel structure map, $\mathcal{F}_{y,x},$ defined in Eq. \ref{defFyx}.    The scaling factor is obtained by first factoring $\mathcal{F}_{y,x}$ into two maps as in \begin{equation}\label{FZW}\mathcal{F}_{y,x}=Z^{y}_{r}W^{r}_{x}.\end{equation}  This gives \begin{equation}\label{RZWR}\bar{C}_{y}=Z^{y}_{r}\bar{C}^{r}_{x} =Z^{y}_{r}W^{r}_{x}\bar{C}_{x}.\end{equation} Here $\bar{C}^{r}_{x}$ is the representation of $\bar{C}_{y}$ on $\bar{C}_{x}.$ $\bar{C}^{r}_{x}$ is scaled, relative to $\bar{C}_{x}$, by the factor $r$.

            This can be expressed by explicit definition of $W^{r}_{x}$ as\begin{equation}\label{Wrx}\begin{array}{c}W^{r}_{x}(a_{x})= ra_{x},\;\;\;\; W^{r}_{x}(\pm_{x})=\pm_{x},\\\\W^{r}_{x}(\times_{x}) =\frac{\mbox{$\times_{x}$}}{\mbox{$r$}},\;\;\;\; W^{r}_{x}(\div_{x})=r\div_{x}.\end{array}\end{equation} The operator $Z^{y}_{r}$,    which takes $\bar{C}^{r}_{x}$ onto $\bar{C}_{y}$, can be defined similarly.

            These equations show that $\bar{C}^{r}_{x}$ has the structure shown in Eq. \ref{CrC} as \begin{equation}\label{Crx}\bar{C}^{r}_{x}=\{C_{x},+_{x},-_{x}, \frac{\times_{x}}{r}, r\div_{x},0_{x},r_{x}\}.\end{equation}Here $r=r_{y,x}$ is the real scaling factor defined on the link from $x$ to $y.$ Also $r_{y,x}$ is a number value in $\bar{C}_{x}.$

            Let $a_{y}$ be a number value in $\bar{C}_{y}.$ \begin{equation}\label{Fxyyx}
             a_{x}=\mathcal{F}_{x,y}a_{y}=\mathcal{F}^{-1}_{y,x}a_{y} =(W^{r}_{x})^{-1}(Z^{y}_{r})^{-1}a_{y}
             \end{equation} is the same number value in $\bar{C}_{x}$ as $a_{y}$ is in $\bar{C}_{y}.$ The scaled representation of $a_{y}$ in $\bar{C}_{x}$ is given by\begin{equation}\label{ryxax}r_{y,x}a_{x}=W^{r}_{x}a_{x} =(Z^{y}_{r})^{-1}a_{y}.\end{equation}

             Scaling in to opposite direction gives a scaled representation of $a_{x}$ on $\bar{C}_{y}.$  This is described by a similar factorization of $\mathcal{F}_{y,x}=\mathcal{F}_{x,y}^{-1}.$

             If $y=x+\hat{\mu}dx$ is a neighbor point of $x,$ then  $r_{y,x}$ is defined by Eq. \ref{defryx}. If $y$ is distant from $x$, then $r_{y,x}$ is defined by Eq. \ref{rgdth} . This assumes that  $\vec{A}(x),$ is the gradient\footnote{The more general case with $\vec{A}$ not the gradient of a scalar field gives $r_[y,x]$ as a path integral of $\vec{A},$ \cite{BenLAM}.} of a  scalar field $\theta(x),$ as shown in Eq. \ref{Agrth}.

             \section{Quantum Theory}\label{QT}
             The effects of local availability of mathematics and  scaling on the quantum mechanical properties of systems need to be investigated. This is especially the case for  quantum properties of systems that are expressed in terms of  integrals and derivatives in space or space time.  To keep things simple, the discussion of these effects, which is based on that in \cite{BenLAM}, will be restricted to non relativistic quantum theory on $3$ dimensional Euclidean space.

             A good example is the description of the wave packet state of a system in state $\psi.$ This has the form
             \begin{equation}\label{psiinty}\psi=\int\psi(y)|y\rangle dy.\end{equation} The complex number, $\psi(y),$  is the amplitude for finding the system at $y$, $|y\rangle$ is a basis state of the system at point $y,$ and the integral is over all space.

             The usual mathematical setting for this description is that $\psi$ is a state in a Hilbert space $H$ with $\psi(y)$ a complex number in $\bar{C}.$ This setting is not appropriate here under the assumption of "mathematics is local" and number scaling. The reason is that $\bar{H}$ and $\bar{C}$ do not belong to any mathematical universes, $\bigcup_{x}$. In this sense they are universal or global spaces and structures. This causes problems in that it is not clear how an observer, at $x,$ whose universe for mathematics is $\bigcup_{x},$ can describe $\psi$ in $\bar{H}$ and $\bar{C}.$

             One approach to this problem is to replace $\bar{H},\bar{C}$ with separate Hilbert space and complex number structures, $\bar{H}_{y},\bar{C}_{y}$ for each point $y$.  This option fits in with the notion of separate mathematical universes with $\bar{H}_{y},\bar{C}_{y}$ belonging to $\bigcup_{y}$ for each $y.$

             This approach is different from that used in gauge theories \cite{Montvay}
             in that, in gauge theories, the vector spaces, $\bar{V}_{y},$ at each point $y$ are finite dimensional. They are spaces for the internal states of matter fields.  Here the $\bar{H}_{y}$ are infinite dimensional.  They can be used to describe basis states over all points in space.  For example, $|w\rangle_{y}$ is the $\bar{H}_{y}$ basis state for a particle at location $w$, $\psi(w)_{y}$ is the amplitude, as a number in $\bar{C}_{y}$, for finding a quantum system in state $\psi$ at $w.$ The resulting wave packet expansion of $\psi_{y} $ is \begin{equation}\label{psiy}\psi_{y}=\int_{y}\psi(w)_{y} |w\rangle_{y} dw_{y}.\end{equation} The subscripts $y$ indicate  that the factors in the integrand and the integral are all in $\bar{H}_{y},\bar{C}_{y}.$

             Scaling has two components. One is external  and the other is internal. External scaling describes the representations, at $x$ of  quantum  states or quantum properties in general of systems at $y.$ Internal scaling applies only to properties of systems expressed as space integrals or space derivatives.

             \subsection{External scaling}\label{ES}

             External scaling is best described by the use of examples.   Let $\psi_{y}$ be a wave packet of a system. The subscript $y$ indicates that it is described in $\bar{H}_{y},\bar{C}_{y}$ as in Eq, \ref{psiy}. The description of $\psi_{y}$ at $x$ is based on vectors and scalars in $\bar{H}_{x}$ and $\bar{C}_{x}.$ With no scaling the description at $x$ is given by \begin{equation}\label{psix}\begin{array}{l}\psi_{x}=\int_{x}\psi(w)_{x} |w\rangle_{x} dw_{x}=U_{x,y}\psi_{y}\\\\\hspace{1cm}=\int_{x} \mathcal{F}_{x,y}(\psi(w)_{y}) U_{x,y}(|w\rangle_{y})\mathcal{F}_{x,y} (dw_{y}).\end{array}\end{equation} Here $U_{x,y}$ is a unitary operator that parallel transforms $\psi_{y}$ to the state $\psi_{x}$ where $\psi_{x}$ is the same state in $\bar{H}_{x}$ as $\psi_{y}$ is in $\bar{H}_{y}.$ $\mathcal{F}_{x,y}$ is defined by Eq. \ref{defFyx} for number structures.

             With scaling included, $\psi_{x}$ is replaced by \begin{equation} \label{psirx} \psi^{r}_{x}=r_{y,x}U_{x,y}\psi_{y}=r_{y,x}\int_{x}\psi(w)_{x} |w\rangle_{x} dw_{x}.\end{equation}Here $\psi_{x}$ is the scaled representation, in $\bar{H}_{x}$ of $\psi_{y}$ in $\bar{H}_{y}.$

             As another example, let \begin{equation}\label{psiylpsiy} \langle\psi_{y}\tilde{l}_{y}\psi_{y}\rangle_{y}=\int_{y}\psi(w)^{*}w_{y} \psi(w)_{y}dw_{y}\end{equation} be the expectation value, at $y,$ for the position of $\psi.$  Here $\tilde{l}_{y}$ is the $\bar{H}_{y}$ position operator at $y.$ With no scaling, the expectation value at $x$ that is the same number in $\bar{C}_{x}$ as $ \langle\psi_{y}\tilde{l}_{y}\psi_{y}\rangle_{y}$ is at $y$ is given by \begin{equation}\label{psixlpsix}
             \langle\psi_{x}\tilde{l}_{x}\psi_{x}\rangle_{x} =\mathcal{F}_{x,y}\int_{y}\psi(w)^{*}w_{y} \psi(w)_{y}dw_{y}=\int_{x} \psi_{x}^{*}(w_{x})w_{x}\psi_{x}(w_{x})dw_{x}. \end{equation}

             With scaling included, the position expectation value at $x$ is given by \begin{equation}\label{psixlpsixr}
             \langle\psi^{r}_{x}\tilde{l}^{r}_{x}\psi^{r}_{x}\rangle^{r}_{x} =r_{y,x}\mathcal{F}_{x,y}\int_{y}\psi(w)^{*}w_{y} \psi(w)_{y}dw_{y}=r_{y,x}\langle\psi_{x}\tilde{l}_{x}\psi_{x}\rangle_{x}. \end{equation} Note that all but one of the $r$ factors  that result from use of  Eq. \ref{psiScx} and $\tilde{l}_{x}^{r}=r_{y,x}\tilde{l}_{x}$ in Eq. \ref{psixlpsixr} are canceled by the scaling of the multiplication or product operations that appear.

             It is good to summarize these results.  Observers, $O_{y}$ and $O_{x}$ at $y$ and $x$ determine the expectation values of the position operator on $\psi$ to be $\langle\psi_{y}\tilde{l}_{y}\psi_{y}\rangle_{y}$ and  $\langle\psi_{x}\tilde{l}_{x}\psi_{x}\rangle_{x}$ respectively. $\langle\psi_{y}\tilde{l}_{y}\psi_{y}\rangle_{y}$ is the same number in $\bar{C}_{y}$ as $\langle\psi_{x}\tilde{l}_{x}\psi_{x}\rangle_{x}$ is in $\bar{C}_{x}.$ With scaling included, the value of $\langle\psi_{y}\tilde{l}_{y}\psi_{y}\rangle_{y},$ as determined by $O_{x}$ at $x,$ is $r_{y,x}\langle\psi_{x}\tilde{l}_{x}\psi_{x}\rangle_{x}.$ This differs by the factor $r_{y,x}$ from the expectation value calculated directly by $O_{x}$.

             The difference between the two values at $x$ is a problem. It will be seen later that this problem can be fixed by placing local restrictions on $r_{y,x}.$

             As noted, external scaling applies to other properties besides spatial ones.  For instance the  momentum operator $\tilde{k}_{y}=ih_{y}(d/dy)_{y}$ corresponds to the scaled momentum operator, \begin{equation}\label{rzxk}r_{y,x}\tilde{k}_{x} =r_{y,x}ih_{x}(d/dw)_{x}\end{equation} at $x.$ Note that the expression for the ith component of the derivative, $d^{i}\psi_{y}(w)/d^{i}w=(\psi_{w} (w+d^{i}w)-\psi_{y}(w))/d^{i}w$ is valid as both amplitudes, $\psi_{y}(w+d^{i}w)$ and $\psi_{y}(w),$ are in $\bar{C}_{y}.$ This is not the case for internal scaling.

             \subsection{Internal scaling}\label{IS}
             In external scaling the integrands of the space integrals in Eqs. \ref{psix} and \ref{psixlpsix} are vectors in $\bar{H}_{x}$.  The integrals are limits of sums of vectors for different space positions.They are elements of $\bigcup_{x}.$ The corresponding integrals at some other location, $y$ are integrals over vectors in $y$. They are elements of $\bigcup_{y}.$ Local availability of mathematics and scaling  applies in transforming the integrals at $ y$ to $x$ (or vice versa).

             In internal scaling local availability of mathematics and scaling is moved inside the integrals.\footnote{This is the reason that the superscripts,  $r$, as in Eq. \ref{psixlpsixr}, are replaced in this section  with $Sc.$. use of $r$ is not suitable here as it depends on the integration space variable.} For each space point $y$ the integrand in the wave packet stat $\psi$ in Eq. \ref{psiy} is a vector in $\bar{H}_{y}$. The integral makes no sense as it is the limit of a sum of vectors in different Hilbert spaces, one for each $y$.  Vector addition is defined only within a Hilbert space.  It is not defined between spaces.

             This can be fixed by transferring each integrand to a point $x$ so that, for each $y,$ the transferred integrand is a vector in $\bar{H}_{x}.$  For the state $\psi$  and with no scaling, $\psi$ becomes
             \begin{equation}\label{psiScx}\begin{array}{l}\psi^{Sc}_{x}=\int_{x} U_{x,y}(\psi_{y}(y)\cdot_{y}|y\rangle_{y}dy)=\int_{x}\mathcal{F}_{x,y} (\psi_{y}(y))\cdot_{x}U_{x,y}(|y\rangle_{y})dy_{x} \\\\\hspace{1cm} =\int_{x}\psi_{x}(y_{x})\cdot_{x}|y_{x}\rangle_{x}dy_{x}.\end{array}\end{equation} This shows that, with no scaling, external and internal scaling coincide, and Eqs. \ref{psiScx} and \ref{psix} are the same expressions for $\psi_{x}.$

             If scaling is used, then Eq. \ref{psiScx} becomes\begin{equation} \label{psiScxr} \begin{array}{l}\psi^{Sc}_{x}=\int_{x} r_{y,x}U_{x,y}(\psi_{y}(y)\cdot_{y}|y\rangle_{y}dy)\\\\\hspace{1cm}=
             \int_{x}r_{y,x}\mathcal{F}_{x,y} (\psi_{y}(y))\cdot^{r}_{x}r_{y,x} U_{x,y}(|y\rangle_{y})dy_{x} \\\\\hspace{2cm} =\int_{x}r_{y,x} \psi_{x}(y_{x})\cdot_{x}|y_{x}\rangle_{x}dy_{x}.\end{array}\end{equation} The scalar-vector multiplications in the different  Hilbert space structures are shown. Here $\cdot^{r}_{x}$ is the scalar vector multiplication in $\bar{H}^{r}_{x},$ which is the scaled representation of $\bar{H}_{y}$ on $\bar{H}_{x}.$ One of the two $r_{y,x}$ factors in the middle integral is canceled by the $r$ factor arising from replacing $\cdot^{r}_{x}$ with $\cdot_{x}.$  Details of this and other properties of $\bar{H}^{r}_{x}$ and $U_{x,y}$ are given in the Appendix.

             The representation, at $x$ of the position expectation value with internal scaling is arrived at in a similar way.  The integral in
             \begin{equation}\label{psilpsi}\langle\psi\tilde{l}\psi\rangle=
             \int\psi^{*}(y)\times_{y}y\times_{y}\psi(y)dy\end{equation} makes no sense because addition is defined only within complex number structures, not between these structures. This is fixed by transferring the integrands, $\psi^{*}(y)\times_{y}y\times_{y}\psi(y)dy,$ in the different $\bar{C}_{y}$ to a reference location, $x$. With scaling this gives, \begin{equation}\label{psilpsir}\begin{array}{l}\langle\psi^{Sc}_{x} \tilde{l}^{Sc}_{x}\psi^{Sc}_{x}\rangle^{Sc}_{x}\\\\\hspace{1cm}=
             \int_{x}r_{y,x}\mathcal{F}_{x,y}(\psi^{*}(y))
             \times^{r}_{x}r_{y,x}\mathcal{F}_{x,y}(y)
             \times^{r}_{x}r_{y,x}\mathcal{F}_{y,x}(\psi(y))dy_{x}\\\\\hspace{2cm}
             =\int_{x}r_{y,x}\psi_{x}(y_{x})\times_{x}y_{x}
             \times_{x}\psi_{x}(y_{x})dy_{x}.\end{array}\end{equation} The relations between the different $\bar{C}_{y}$ and their scaled representations on $\bar{C}_{x}$ are given in Section \ref{NS}.

\begin{figure}[h!]\begin{center}
            \rotatebox{270}{\resizebox{160pt}{160pt}{\includegraphics[80pt,130pt]
            [490pt,540pt]{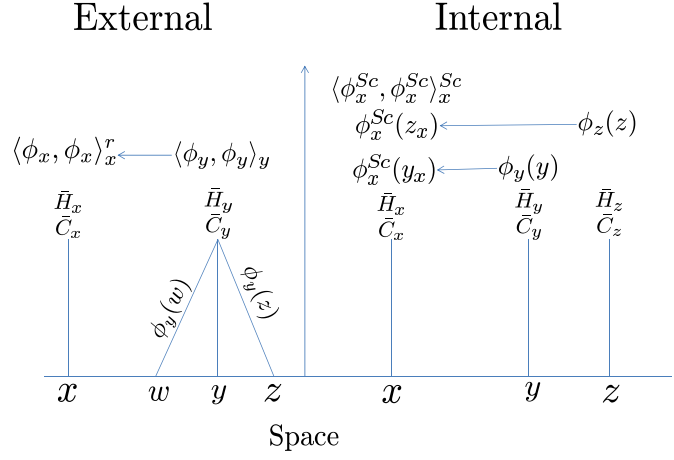}}}\end{center}
            \caption{Comparison of the effects or external and internal scaling on $\langle\phi,\phi\rangle.$ For external scaling the value of $\langle\phi,\phi\rangle,$ at $y$ is expressed as a space integral in $\bar{H}_{y},\bar{C}_{y}$. This shown by the three lines from different space points, $w,y,z$ all converging to $\bar{H}_{y},\bar{C}_{y}.$ For internal scaling, the value of the integrand at each space point is in the Hilbert space complex number structures associated with the point.  This is shown by the two vertical lines at $y,z.$ The integrands are then mapped to corresponding integrands, with scaling, to a common point,  $x,$ where the  integral is defined on $\bar{H}_{x},\bar{C}_{x}.$}\label{SPIE22}\end{figure}

             The relationship between external and internal scaling is shown schematically in Figure \ref{SPIE22} for the norm $\langle\phi,\phi\rangle$ of $\phi.$ The distinction between vectors $\phi(y)|y\rangle$ all in $\bar{H}_{x}$ as in external scaling and in separate $\bar{H}_{y}$ as in internal scaling is clearly shown. Carrying out the integration $\int\int dydy'\phi^{*}(y)\langle y|y'\rangle\phi(y')=\int \phi^{*}(y)\phi(y)dy$ converts the integrals from integration over the  $\bar{H}_{y}$ to integration over the  $\bar{C}_{y}$ as was discussed in the text.\\

             Internal scaling includes external scaling.  To see this, one notes that transfer, with scaling, of $\langle\psi^{Sc}_{x} \tilde{l}^{Sc}_{x}\psi^{Sc}_{x}\rangle^{Sc}_{x}$ to another point, $z$, gives \begin{equation}\label{psizwx}\langle\psi^{Sc}_{z} \tilde{l}^{Sc}_{z}\psi^{Sc}_{z}\rangle^{Sc}_{z} =r_{x,z}\mathcal{F}_{z,x}\langle\psi^{Sc}_{x} \tilde{l}^{Sc}_{x}\psi^{Sc}_{x}\rangle^{Sc}_{x}. \end{equation} One can move the $r_{x,z}$ factor inside the integral as $(r_{x,z})_{x}$ (the same number in $\bar{C}_{x}$ as $r_{x,z}$ is in $\bar{C}_{z}$) and use $r_{y,z}=r_{y,x}(r_{x,z})_{x}$ and remove the $x$ subscripts. This gives  Eq. \ref{psilpsir} with $z$ replacing $x$ everywhere.

             A convenient approach to the use of scaling in quantum mechanical expressions is to note that the usual basis expansion in space states, $1=\int|y\rangle\langle y|dy$  is replaced by\begin{equation}\label{1xapp} 1_{x}\approx  \int_{x}r_{y,x}U_{x,y}|y\rangle\langle y|U_{y,x}dy_{x}= \int_{x}r_{y,x}|y_{x}\rangle_{x}\langle y_{x}|dy_{x}.\end{equation} Here $U_{y,x}=U^{\dag}_{x,y}.$ Justification for the use of Eq. \ref{1xapp} will be discussed shortly.

             An example of the use of Eq. \ref{1xapp} is the relation between the momentum eigenstate $|k\rangle_{x}$  and position eigenstates at $x.$ For example, \begin{equation}\label{kx} |k_{x}\rangle_{x}=\int_{x}r_{y,x} |y_{x}\rangle_{x}\langle y_{x}|k_{x}\rangle_{x}dy_{x}=\int_{x}r_{y,x} e^{ik_{x}\cdot_{x}y_{x}}|y_{x}\rangle_{x} dy_{x}.\end{equation} The momentum operator $\tilde{k}_{x}$ satisfies the usual relations such as  \begin{equation}\label{tilkx}
             \tilde{k}_{x}\psi_{x}=\int_{x}\tilde{k}_{x}|k_{x}\rangle_{x}\langle k_{x}|\psi\rangle_{x}dk_{x} =\int_{x}k_{x}|k_{x}\rangle_{x}\psi(k)_{x}dk_{x}.\end{equation}

             Scaling introduces the components, $A_{j}(x)$, of the gradient of the scalar field into the momentum. This can be seen from the expression $\vec{p}=ihd/dy.$ At $y$ the action of  the $jth$ component of $\vec{p}_{y}$ on $\psi$ is given by\begin{equation}\label{mompi}\begin{array}{l}p^{j}_{y}\psi_{y}(y) =i_{y}h_{y}\frac{\textstyle r_{y+d^{j}y,y}\mathcal{F}_{y,y+d^{j}y}\psi(y+d^{j}y)-\psi(y)}{\textstyle d^{j}y}\\\\\hspace{1cm}=i_{y}h_{y}\frac{\textstyle r_{y+d^{j}y,y}\psi(y+d^{j}y)_{y}-\psi(y)}{\textstyle d^{j}y}.\end{array}\end{equation} The factor $r_{y+d^{j}y,y}\mathcal{F}_{y,y+d^{j}y}$ accounts  for the scaling arising from the transforming $\psi(y+d^{j}y),$ which belongs to $\bar{C}_{y+d^{j}y},$ to an element of $\bar{C}_{y}$ on which the derivative is defined.

             Following the procedure used in gauge theories \cite{Cheng} in deriving covariant derivatives, one expands $r_{y+d^{j}y,y}$ to first order to obtain  $r_{y+d^{j}y,y}=e^{A_{j}(y)d^{j}y}\approx 1_{y}+A_{j}(y)d^{j}y.$ This is used in Eq. \ref{mompi} to obtain \begin{equation}\label{mompiD}p^{j}_{y}\psi_{y}(y) =i_{y}h_{y}(\frac{d^{\prime,j}}{d^{j}y}+A_{j}(y))\psi(y). \end{equation} This shows that, in the presence of scaling, the momentum, at $y$,  $\vec{p}_{y},$ has the canonical form,\begin{equation}\label{mompy} \vec{p}_{y}=i_{y}h_{y}(\frac{d^{\prime}}{dy}+\vec{A}(y)).\end{equation} The prime on $d^{\prime}/dy$ accounts for the local availability of mathematics in that $\psi(y+d^{j}y)_{y}$ is the same value in $\bar{C}_{y}$ as $\psi(y+d^{j}y)$ is in $\bar{C}_{y+d^{j}y}.$ It has no effect on the value of the derivative. Also, Eq. \ref{Agrth}, $\vec{A}(y)=\nabla_{y}\theta.$

             There is another important distinction to be made between external and internal scaling. physical laws expressed as equations, and equations in general, are invariant under external scaling. This is the case even though the individual terms of the equations are not invariant. This is a consequence of the cancelation of all but one of the scaling factors in the terms on each side of the equation. As an example, consider the Einstein mass energy relation, $E=mc^{2}.$ At $y$ this relation is $E_{y}=m_{y}c_{y}^{2}$.  Here $c_{y}$ is the number value in $\bar{C}_{y}$ for the velocity of light. This relation corresponds to the scaled relation \begin{equation}\label{Erx}
             E^{r}_{x}=m^{r}_{x}\times^{r}_{x}c^{r}_{x}\times^{r}_{x}c^{r}_{x}
             \end{equation} in $\bar{C}^{r}_{x}.$  Here $r=r_{y,x}.$ Replacement of $E^{r}_{x},m^{r}_{x},c^{r}_{x}$ and $\times^{r}_{x}$ by their scaled values, $rE_{x},rm_{x},rc_{x},$ and $\times/r$ in $\bar{C}_{x}$ gives $rE_{x}=rm_{x}c^{2}_{x}$, or $E_{x}=m_{x}c_{x}^{2}.$ This is the same equation in $\bar{C}_{x}$ as $E_{y}=m_{y}c_{y}^{2}$ is in $\bar{C}_{y}.$

             However, internal scaling is preserved in  equations  under scaled transformations from $y$ to $x.$ It does not cancel out under transferral as does external scaling. As an example, consider the well known quantum equation $\tilde{H}\psi=E\psi$ for the energy of a system in state $\psi.$  $\tilde{H}$ is the Hamiltonian for the system. At location $y$, expansion in a complete set of position states, using Eq. \ref{1xapp}, gives
             \begin{equation}\label{HpsiE}\int_{y}r_{w,y}\tilde{H}_{y}|w_{y} \rangle_{y}\psi_{y}(w_{y})dw_{y}=E_{y}\psi_{y}.\end{equation}Transferral to site $x,$ with external scaling, gives $$r_{y,x}U_{x,y}\int_{y}r_{w,y}\tilde{H}_{y}|w_{y} \rangle_{y}\psi_{y}(w_{y})dw_{y}=r_{y,x}U_{x,y}(E_{y}\psi_{y}).$$ canceling the $r_{y,x}$ factors on both sides of the equation and acting with $U_{x,y}$ gives,
             \begin{equation}\label{rHx}\int_{x}(r_{w,y})_{x}\tilde{H}_{x}|w_{x} \rangle_{x}\psi_{x}(w_{x})dw_{x}=E_{x}\psi_{x}.\end{equation}Here $(r_{w,y})_{x}$ is the same number in $\bar{C}_{x}$ as $r_{w,y}$ is in $\bar{C}_{y}.$

             Note that no scaling factor appears if one expands in eigenstates of $\tilde{H}$ instead of position states. The apparent inconsistency between these two basis expansions is another reason why the discussion in the next section of the values of the scaling factor is needed.

             \section{The value of $r_{y,x}$}\label{Vr}
             The results for quantum mechanics shown so far  suggest that $r_{y,x}\approx 1_{x}.$  This needs to be investigated in more detail.  To begin, one notes that all wave packets have finite spatial extension. Let $Z$ be a finite spatial region that includes essentially all of $\psi,$ and $W$ the rest of space. Then expectation values, at $x$ with scaling included, for the position of a system have the form \begin{equation}\label{pospsi}
             \langle \psi^{Sc}_{x},\tilde{l}^{Sc}\psi^{Sc}_{x}\rangle^{Sc}_{x} \approx\int_{Z}r_{y,x}\psi(y)_{x}^{*} y_{x}\psi(y)_{x}dy_{x}. \end{equation} For momentum expectation values, $\tilde{l}$ is replaced by the momentum operator given by Eq. \ref{mompy}.

             It is useful to separate $r_{y,x}$ into components, external scaling and internal scaling.  Let $z$ be a point on or near the surface of $Z.$ Then using $r_{y,x}=(r_{y,z})_{x}r_{z,x},$ one has \begin{equation}\label{pospsixzy}\langle \psi^{Sc}_{x},\tilde{l}^{Sc}\psi^{Sc}_{x} \rangle^{Sc}_{x} \approx r_{z,x}\int_{Z}(r_{y,z})_{x}\psi(y)_{x}^{*} y_{x}\psi(y)_{x}dy_{x}. \end{equation} As noted this is the expectation value for an observer, $O_{x}.$

             For an observer, $O_{z}$ at $z$, the expectation value, as a number in $\bar{C}_{z},$ is \begin{equation}\label{pospsi}\langle\psi^{Sc}_{z},\tilde{l}^{Sc}\psi^{Sc}_{z} \rangle^{Sc}_{z} \approx\int_{Z}r_{y,z}\psi(y)_{z}^{*}y_{z}\psi(y)_{z}dy_{z}. \end{equation}

             For wave packet states encountered in quantum mechanics, the region, $Z,$ is  small. It follows from this that the distance over which $r_{y,z}$ extends is small. In this case, one  requires that, for all $y$ in $Z,$  \begin{equation}\label{ryzth} r_{y,z}= e^{\theta(y)-\theta(z)}\approx 1_{z}.
             \end{equation} To first order in $\theta,$ this gives\begin{equation}\label{thyz}\theta(y)-\theta(z)
             \approx 0.\end{equation}That is, $\theta$ is roughly constant over $Z.$

             The situation is different for an observer $O_{x}$ at a point $x$  that can be  far away from $Z.$  To see this, recall that, for  observer $O_{z}$ at $z,$ the position expectation  value is $\langle\psi_{z}\tilde{l}_{z}\psi_{z} \rangle_{z} =\int_{z}\psi^{*}(y_{z})y_{z}\psi_{z}(y)dy_{z}.$ For $O_{x}$ at $x$, the value is $\langle\psi_{x}\tilde{l}_{x}\psi_{x}\rangle_{x} =\int_{x}\psi^{*}(y_{x})y_{x}\psi_{x}(y)dy_{x}.$  Internal scaling has been ignored here, for the reasons noted above.  Furthermore $\langle\psi_{x}\tilde{l}_{x}\psi_{x}\rangle_{x}$ is the same number value in $\bar{C}_{x}$ as $\langle\psi_{z}\tilde{l}_{z}\psi_{z}\rangle_{z}$ is in $\bar{C}_{z}$.  This follows from\begin{equation}\label{exppsilxFz} \langle\psi_{x}\tilde{l}_{x}\psi_{x}\rangle_{x}=
             \mathcal{F}_{x,z}\langle\psi_{z}\tilde{l}_{z}\psi_{z}\rangle_{z}.\end{equation}

             However, with external scaling included, the expectation value at $z$ corresponds to the value\begin{equation}\label{rzxexpp} r_{z,x}\mathcal{F}_{x,z} \langle\psi_{z}\tilde{l}_{z}\psi_{z}\rangle_{z} =r_{z,x}\langle\psi_{x}\tilde{l}_{x}\psi_{x}\rangle_{x}.\end{equation} This is $O_{x}'s$ representation of the expectation value for $O_{z}$ at $z.$

              $O_{x}$ now has a quandry: which expectation value is correct, the one with $r_{z,x}$ or the one without. This can be resolved by  $O_{z}$ actually measuring the expectation value at or near $Z.$

              After carrying out the measurement, $O_{z}$ finds that, to within experimental error, the experimental result agrees with his expected value, $\langle\psi_{z}\tilde{l}_{z}\psi_{z}\rangle_{z}$ as determined by calculation. $O_{z}$ transmits the experimental result to $O_{x}$ so $O_{x}$ can compare the result with his/her calculation.

             As was shown in Section \ref{LAM},  scaling is not involved in physical transmission of results of calculations or experiment. Thus $O_{x}$ will interpret the transmitted signal to be the same number as was found by $O_{z}.$ $O_{x}$  will then conclude from this that the scaled expectation value in Eq. \ref{rzxexpp} agrees with experiment  only if  $r_{z,x}\approx 1_{x},$ or $\theta(z)-\theta(x)\approx 0$ to experimental accuracy.

             There are a great many regions $Z$, which can be used to represent  expectation values,  carry out experiments, and computations.  Let $\bigcup Z$ represent the union of all these regions. The region $\bigcup Z$  is quite large. It contains all  space regions in which observers can prepare and observe states of quantum systems that are wave packets or, in the case of multiple systems, are entangled states, as in quantum teleportation \cite{Bennett,Bennett2}.

             The detailed discussion shows that $\theta(z)-\theta(x)\approx 0$ for all pairs of points, $x,z$ in $\bigcup Z.$ In fact one can set $\theta(z)-\theta(x)\approx 0$ in a larger region, $L,$ that contains $\bigcup Z.$ The region, $L,$ should be such that observers at different locations in $L$, can, in principle, prepare systems in quantum states,  make measurements on them, and efficiently communicate details and results of measurements and calculations to one another.

             The region $L$ is large. It includes all terrestrial locations,  probably all locations in the solar system, and  possibly more outside.  A generous estimate of the size of $L$ is as a sphere with a radius of several light years centered roughly on the sun. This is based on the use of the value of about $1,000$ light years \cite{AmSci} as the limit of the size of the space region in which we, as terrestrial observers, can detect intelligent life on other worlds. The reason that $L$ is smaller that $1000$ light years is that multiple transmissions of information between us and a distant world are needed to have distant beings carry out experiments and transmit the results to us. Also we have to be sure that the transmitted results are for experiments we think they are for. The exact size of $L$ is not important. However it should be a very small fraction of the total size of the universe.

             \subsection{Effects of scaling outside of $L$.}\label{ESOL}

             The reasons described above  for setting $r_{z,x}\approx 1_{x},$ fail for regions $Z$ outside $L.$ The reason is that communications between observers, if any, outside $L$ and those inside take too long to be useful.  Thus there is no way for observer $O_{x}$ at $x$ inside $L$, to tell whether the scaled representation, $r_{z,x}\langle\psi_{x}\tilde{l}_{x}\psi_{x}\rangle_{x},$ at $x,$ of the expectation value at $z$ outside $L$ is correct or not. Neither calculated results nor experimental results can be effectively transmitted from $z$ to $x$ inside $L.$ Also the time required for multiple transmissions of information from $x$ to $z$ and back so $O_{x}$ can obtain the details of the experiment and calculation to be sure that the correct experiment or calculation was done, would take too long.

             It follows that for points $x$ in $L$ and $z$ outside $L,$ $r_{z,x}\neq 1_{x}$ or $\theta(z)-\theta(x)\neq 0$ must be considered. In particular this suggests that number scaling, as represented by the scalar field, $\theta,$ may be important in cosmology.\footnote{Observations of properties of cosmological systems are not experiments in that there are no state preparations by observers. One is limited to measurement of the properties of incoming photons and particles.}  However this is a subject for future work.

             \section{Discussion}\label{D}
             There is much work to be done to further develop  the basic ideas of local availability of mathematics and number scaling.  Here, and in other work \cite{BenLAM} the effects of these ideas on some simple aspects of quantum theory were investigated.  The work needs to be expanded into areas of quantum theory other than those discussed here. Also the restriction to  $3$ dimensional Euclidean space needs to be relaxed to include $4$
             dimensional space time as in relativity theory. Also the effects on geometry need investigation.

             However, in spite of the limited nature of this work, it was possible to draw some conclusions.  One was the separation of scaling into external and internal components. External scaling applies to all physical quantities as number values when viewed at different locations.  If $a_{z}$ is a physical quantity as a number value at $z$, then this value corresponds to the scaled  number value $a^{Sc}_{x}=r_{z,x}\mathcal{F}_{x,z}a_{x}=r_{z,x}(a_{z})_{x}$ at $x$.  This is the representation, with external scaling included, of the number value, $a_{z},$ at $x.$  Note that $(a_{z})_{x}$ is the same number value in $\bar{C}_{x}$ as $a_{z}$  is in $\bar{C}_{z}.$

             Wave packet states and expectation values as integral over space need separate consideration. If one assumes the existence at each point, $x,$ a Hilbert space structure $\bar{H}_{x}$ that can be used to express wave packets and expectation values as integrals over position vectors in $\bar{H}_{x},$ such as in Eqs.  \ref{psix} and \ref{psiylpsiy}, then external scaling is sufficient.

             However, if one assumes that local availability of mathematics apples separately to each position vector $|y\rangle,$ then the  space integrals,  as in Eqs.  \ref{psix} and \ref{psiylpsiy} do not make sense. The reason is that they are integrals over vectors in different Hilbert space structures. This remedied be transferring the integrands at each $y$ to a common point $x$ at which the integral do make sense. Internal scaling refers to the scaling, along with the parallel transferring, of quantities at $y$ to $x$ for integration. In this case the scaling factor $r_{y,x}$ depends on the integration variable, $y.$  Eqs. \ref{psiScxr} and \ref{psilpsir} show the effect of internal scaling.

             The large amount of experimental support for quantum mechanics show that, to within the limits of experimental accuracy, no scaling, external or internal, is present.  That is $r_{z,x}=e^{\theta(z)-\theta(x)}\approx 1$ or $\theta(z)-\theta(x)\approx 0.$  It was seen here that this result depends on the ability of an observer at point, $z,$ to transmit the results of experiment to an observer at another point, $x$. This is based on the fact, discussed in section \ref{LAM}, that no scaling is involved in the physical transmission of information from one point to another.

             This shows that that the limitation $r_{z,x}\approx 1$ applies only to regions of space time in which it is possible for observers to carry out experiments and communicate effectively with us as observers  at points $x$ within a region that includes the solar system. If the points $z$ at which observers are to carry out experiments are more than a few light years distant from $x$ it becomes effectively for the distant observers to communicate with us. This is especially so because any experimental check at $z$ of a prediction made at $x$ requires multiple two way communications between observers at $x$ and $z.$

             It follows that the restriction $\theta(z)-\theta(x)\approx 0$ applies only to points within a few light years of one another and with $x$ in a region of a few light years in diameter that includes the solar system.  If $z$ is outside the region, at cosmological distance for example, then $\theta(z)-\theta(x)\neq 0$ is possible as it cannot be refuted by experiments, at least of the type considered here.

             One additional point to note is that the value of $r$ is invariant under changing $\theta$ everywhere by a constant. If $\theta'(x)=\theta(x)+c$, then for all $z,x$ $\theta'(z)-\theta'(x)=\theta(z)-\theta(x).$

             \section*{Acknowledgement}
            This work was supported by the U.S. Department of Energy,
            Office of Nuclear Physics, under Contract No.
            DE-AC02-06CH11357.

            \appendix
            \section{Appendix}

             So far, the operators $\mathcal{F}_{y,x}$ for different $y,x$ and their factorization have been described for relating complex number structures at different locations.  This needs to be extended to Hilbert spaces, especially in view of their incorporation of scalars into their basic properties. To this end, let $\bar{H}_{x}$ in $\bigcup_{x}$ and $\bar{H}_{z}$ in $\bigcup_{z}$ be separable Hilbert space structures defined by Eq. \ref{Hilbert}. Let $U_{x,z}:\bar{H}_{z}\rightarrow\bar{H}_{x}$ be a unitary operator defined by \begin{equation}\label{Uxz}\begin{array}{c}U_{x,z}H_{z}=H_{x}\;\;\;\;U_{x,z}\pm_{z}=\pm_{x},
             \\U_{x,z}(\cdot_{z})=\cdot_{x}\;\;\;\;U_{x,z}\phi_{z}=(\phi_{z})_{x},\\U_{x,z}\langle\phi_{z},\psi_{z} \rangle_{z}=\langle(\phi_{z})_{x},(\psi_{z})_{x} \rangle_{x}=\mathcal{F}_{x,z}(\langle\phi_{z}, \psi_{z}\rangle_{z}).\end{array}\end{equation}$U_{x,z}$ is a parallel transform operator in that $(\phi_{z})_{x}$ and $(\psi_{z})_{x}$ are the same states in $\bar{H}_{x}$ as $\phi_{z}$ and $\psi_{z}$ are in $\bar{H}_{z}.$ Also $\mathcal{F}_{x,z}$ is the parallel transform operator, Eq. \ref{defFyx}, applied to complex number structures. Note too that if $\phi_{z}=a_{z}\cdot_{z}\theta_{z}$ then $U_{x,z}\phi_{z}=\mathcal{F}_{x,z}(a_{z})\cdot_{x}(\theta_{z})_{x}.$ Here $\mathcal{F}_{x,z}(a_{z}) =(a_{z})_{x}$ is the same number in $\bar{C}_{x}$ as $a_{z}$ is in $\bar{C}_{z}.$

             The states, $\phi_{z}$  with $\psi_{x}$ cannot be directly compared as they are in different Hilbert space structures. However, one can use   $U_{x,z}$  to parallel transport $\phi_{z}$ to $\bar{H}_{x}$ for comparison with $\psi_{x}.$ An example of such a comparison is expressed by the norm,  $|(\phi_{z})_{x}-\psi_{x}|_{x}$ in $\bar{H}_{x}$  The comparison can just as easily be done in $\bar{H}_{z}$ by using $U_{z,x}=U^{\dag}_{x,z}$ to parallel transport $\psi_{x}$ to $\bar{H}_{z}.$

             To introduce scaling, one first needs to define a local scaled representation of $\bar{H}_{z}$ on $\bar{H}_{x}$. Following the factorization of $\mathcal{F}_{z,x},$ Eqs. \ref{FZW}-\ref{Wrx}, $U_{z,x}$ is factored to give\begin{equation}\label{UXX}U_{z,x}=X^{z}_{r}X^{r}_{x}\end{equation} where
             \begin{equation}\label{UXXH}\bar{H}_{z}=U_{z,x}\bar{H}_{x}= X^{z}_{r}\bar{H}^{r}_{x}=X^{z}_{r}X^{r}_{x}\bar{H}_{x}.\end{equation}

             $\bar{H}^{r}_{x}$ is the local, scaled representation, in $\bigcup_{x}$, of $\bar{H}_{z}$ on $\bar{H}_{x}.$ It is given \cite{BenIJTP,BenLAM} by   \begin{equation}\label{Hrx}\bar{H}^{r}_{x} =\{H_{x},\pm^{r}_{x},\cdot^{r}_{x},\langle\ -,-\rangle^{r}_{x}, \psi^{r}_{x}\}=\{H_{x},\pm_{x}, \frac{\cdot_{x}}{r},\frac{\langle\ -,-\rangle_{x}}{r},r\psi_{x}\}.\end{equation} The $r=r_{z,x}$ factors appearing with $\cdot_{x}$ and $\langle-,-\rangle_{x}$ are necessary so that $\bar{H}^{r}_{x}$ satisfies the Hilbert space axioms\footnote{Another definition of $\bar{H}^{r}_{x}$ that satisfies the equivalency of axiom satisfaction is $$\{H_{x},\pm_{x},\frac{\cdot_{x}}{r},r\langle\ -,-\rangle_{x},\psi_{x}\}.$$ This definition , which lacks a scaling factor for states but preserves it for amplitudes and scalar products, is not used because it does not satisfy the equivalence between Hilbert spaces and products of complex number structures \cite{Kadison}. The definition used here satisfies the equivalence $\bar{H}^{r}_{x}\cong (\bar{C}^{r}_{x})^{n}$ for finite dimensional spaces.} if and only if $\bar{H}_{x}$ does \cite{Kadison}. Also $\psi^{r}_{x}$ is the same vector in $\bar{H}^{r}_{x}$ as $\psi_{x}$ is in $\bar{H}_{x}.$

              It is interesting to note that $U_{z,x}$ and $X^{z}_{r}$ map Hilbert spaces in $\bigcup_{x}$ to Hilbert spaces in $\bigcup_{z}.$ As such they do not appear to be  elements of either universe. This is supported by the fact that they do not have representations as matrices of numbers or as exponentials of Lie algebra generators.  Only $X^{r}_{x}$ has  these representations.  And it belongs to $\bigcup_{x}.$

              Comparison of properties of $\phi_{z}$ with $\psi_{z}$ at a point $x\neq z$ uses scaling. This is seen by noting that $\phi^{r}_{x}=r_{z,x}U_{x,z}\phi_{z}=r_{z,x}(\phi_{z})_{x}.$ The local scaled representation of $\phi_{z}$ on $\bar{H}_{x}$ is obtained by multiplying the parallel transformation of $\phi_{z}$  to $x$ by the scaling factor, $r_{z,x}.$

            \end{document}